\documentclass[journal]{IEEEtran}

\usepackage[utf8]{inputenc}
\usepackage[T1]{fontenc}
\usepackage{amsmath}
\usepackage{amssymb}
\usepackage{graphicx}
\usepackage{booktabs}
\usepackage{cite}
\usepackage{subcaption}
\usepackage{url}
\usepackage{float}
\usepackage{placeins}

\AtBeginDocument{%
  \abovedisplayskip=10pt plus 3pt minus 3pt
  \belowdisplayskip=10pt plus 3pt minus 3pt
  \abovedisplayshortskip=6pt plus 3pt
  \belowdisplayshortskip=6pt plus 3pt
}

\begin{document}

\title{A Scaling Law for Bandwidth Under Quantization}

\author{Maximilian~Kalcher and Tena~Dubcek%
\thanks{M. Kalcher is with ETH Zurich, Switzerland (e-mail: mkalcher@ethz.ch).}%
\thanks{T. Dubcek is with ETH Zurich and the Swiss Epilepsy Center, Clinic Lengg, Z\"urich, Switzerland (e-mail: dubcekt@ethz.ch).}}

\maketitle

\begin{abstract}
We derive a scaling law relating ADC bit depth to effective bandwidth for signals with $1/f^\alpha$ power spectra. Quantization introduces a flat noise floor whose intersection with the declining signal spectrum defines an effective cutoff frequency $f_c$. We show that each additional bit extends this cutoff by a factor of $2^{2/\alpha}$, approximately doubling bandwidth per bit for $\alpha = 2$. The law requires that quantization noise be approximately white, a condition whose minimum bit depth $N_{\min}$ we show to be $\alpha$-dependent. Validation on synthetic $1/f^\alpha$ signals for $\alpha \in \{1.5, 2.0, 2.5\}$ yields prediction errors below 3\% using the theoretical noise floor $\Delta^2/(6f_s)$, and approximately 14\% when the noise floor is estimated empirically from the quantized signal's spectrum. We illustrate practical implications on real EEG data.
\end{abstract}


\section{Introduction}

The Signal-to-Quantization-Noise Ratio (SQNR) for an $N$-bit uniform quantizer is given by $\text{SQNR} = 6.02N + 1.76$~dB \cite{bennett1948}. This classical result describes total noise power under the assumption that quantization error is white and uniformly distributed, but provides no information about frequency-dependent effects.

Many signals of practical interest exhibit power-law spectra $S(f) \propto 1/f^\alpha$: electrophysiology data such as electroencephalograms (EEG) with $\alpha \approx 1.5$--$2$ \cite{he2010, freeman2006}, audio signals, ocean noise \cite{wenz1962}, and seismic recordings \cite{aki1980}. When such signals undergo amplitude quantization, practitioners commonly observe an apparent low-pass filtering effect, with high-frequency content degraded before low-frequency content. Despite the prevalence of this phenomenon, no formula exists in the literature that predicts the relationship between bit depth and effective bandwidth for power-law spectra \cite{gray1990, widrow2008}.

\begin{figure}[t]
\centering
\begin{subfigure}[b]{0.48\columnwidth}
    \includegraphics[width=\textwidth]{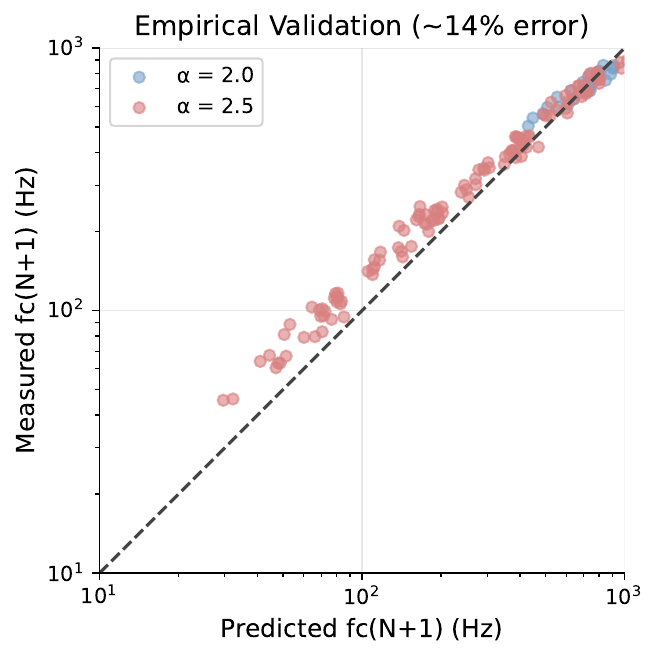}
    \caption{Empirical (${\sim}14\%$ error)}
\end{subfigure}
\hfill
\begin{subfigure}[b]{0.48\columnwidth}
    \includegraphics[width=\textwidth]{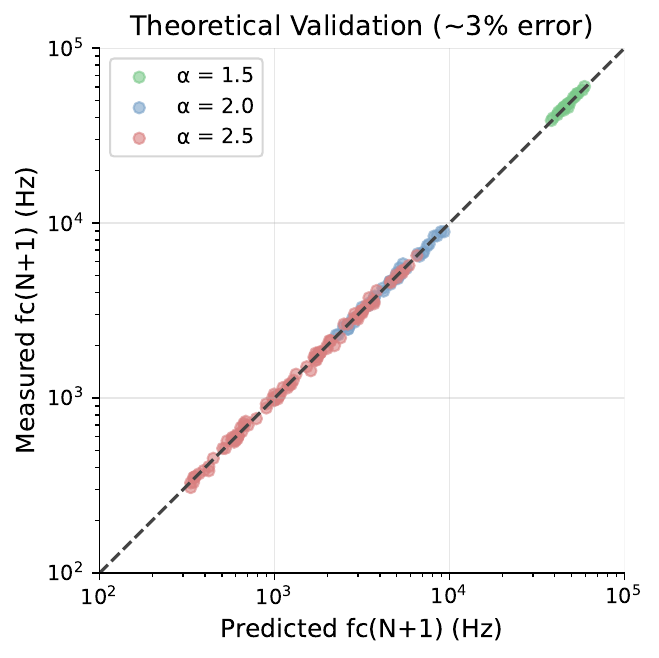}
    \caption{Theoretical (${<}3\%$ error)}
\end{subfigure}
\caption{Scaling law validation: $f_c(N{+}1)$ predicted from Eq.~\eqref{eq:scaling} vs.\ measured values. Empirical noise floor estimation introduces spectral artifacts~(a), while the theoretical noise floor yields close agreement~(b).}
\label{fig:scaling}
\end{figure}

In this paper, we derive a scaling law for this relationship:
\begin{equation}
\boxed{\frac{f_c(N+1)}{f_c(N)} = 2^{2/\alpha}}
\label{eq:scaling}
\end{equation}
where $f_c(N)$ is the effective cutoff frequency at bit depth $N$ and $\alpha$ is the spectral slope. For $\alpha = 2$ (Brownian motion), each additional bit doubles the usable bandwidth; for $\alpha = 1.5$ (typical electrophysiology data), each bit extends bandwidth by approximately $2.5\times$. The key insight is that quantization is \emph{not} frequency-selective; it introduces broadband noise whose flat spectrum intersects the declining $1/f^\alpha$ signal spectrum at a well-defined cutoff frequency $f_c$, above which quantization noise dominates. We further show that the white noise assumption underlying this result requires a minimum bit depth $N_{\min}(\alpha)$ that increases with the spectral slope (Table~\ref{tab:scaling}). Fig.~\ref{fig:scaling} shows experimental validation of the scaling law.

\section{Theory}

\subsection{Signal and Noise Model}

Consider a signal with power spectral density (PSD)
\begin{equation}
S(f) = S_0 \, f^{-\alpha}
\end{equation}
where $S_0$ is the PSD at $f = 1$~Hz and $\alpha > 0$ is the spectral slope. Under the additive white noise model of quantization \cite{bennett1948, sripad1977, widrow2008}, the one-sided noise PSD is
\begin{equation}
N_q = \frac{\Delta^2}{6 f_s}
\end{equation}
where $\Delta = R/2^N$ is the quantization step size, $R$ is the signal range, and $f_s$ is the sampling frequency.

\subsection{Cutoff Frequency}

We define the effective cutoff frequency $f_c$ as the frequency at which the signal PSD equals the quantization noise floor:
\begin{equation}
S_0 \, f_c^{-\alpha} = \frac{\Delta^2}{6 f_s}
\end{equation}
Solving for $f_c$ and substituting $\Delta = R/2^N$ yields the closed-form cutoff frequency:
\begin{equation}
f_c(N) = \left( \frac{6 S_0 f_s}{R^2} \right)^{1/\alpha} \cdot\, 2^{2N/\alpha}
\label{eq:cutoff}
\end{equation}
This expression gives the absolute bandwidth for a given bit depth $N$ directly from the signal parameters: the spectral coefficient $S_0$, sampling frequency $f_s$, ADC range $R$, and spectral slope $\alpha$. The first factor is a signal-dependent constant, while the second factor $2^{2N/\alpha}$ captures the exponential growth of bandwidth with bit depth.

\subsection{Derivation of the Scaling Law}

When comparing bit depths $N$ and $N+1$, the signal-dependent parameters $S_0$ and $f_s$ cancel:
\begin{equation}
\frac{f_c(N+1)}{f_c(N)} = \left( \frac{\Delta_N}{\Delta_{N+1}} \right)^{2/\alpha}.
\end{equation}
Since $\Delta_N / \Delta_{N+1} = 2$, this reduces to
\begin{equation}
\frac{f_c(N+1)}{f_c(N)} = 2^{2/\alpha}.
\end{equation}
The result depends only on the spectral slope $\alpha$ and is independent of $S_0$, $R$, and $f_s$. Table~\ref{tab:scaling} lists the predicted scaling factors.

\begin{table}[ht]
\caption{Bandwidth scaling factor $2^{2/\alpha}$ per additional bit, with empirically determined minimum bit depth $N_{\min}$ for white quantization noise (Section~\ref{sec:noise}).}
\centering
\begin{tabular}{llll}
\toprule
$\alpha$ & Scaling factor & $N_{\min}$ & Example \\
\midrule
1.0 & $4.00\times$ & 4 & Pink noise \\
1.5 & $2.52\times$ & 5 & Electrophysiology \\
2.0 & $2.00\times$ & 7 & Brownian motion \\
2.5 & $1.74\times$ & 10 & Steep spectra \\
\bottomrule
\end{tabular}
\label{tab:scaling}
\end{table}

\subsection{Validity Conditions}

The derivation relies on the white noise model of quantization error, which holds when the signal traverses many quantization levels per sample \cite{sripad1977}. For $1/f^\alpha$ signals, energy concentration at low frequencies means that high-frequency components may span only a few quantization levels at low bit depths, causing the quantization error to become signal-dependent and spectrally colored \cite{jimenez2011}. We determine the threshold $N_{\min}(\alpha)$ empirically in Section~\ref{sec:noise}.

\section{Experiments}

\subsection{Quantization Noise Spectrum}
\label{sec:noise}

We generated $1/f^\alpha$ signals ($10^5$ samples, $f_s = 2000$~Hz, 20 trials) for $\alpha \in \{1.0, 1.5, 2.0, 2.5, 3.0\}$ and measured the spectral slope of the quantization noise $e[n] = x_q[n] - x[n]$ at bit depths 4 through 12 via least-squares linear fit in log-log coordinates.

Table~\ref{tab:noise} shows results for $\alpha = 2$. At 4 bits the noise spectrum has slope $-1.16$, indicating that the error becomes spectrally colored. The slope decreases in magnitude with increasing bit depth and reaches approximately zero at 7 bits, confirming the transition to white noise.

\begin{table}[t]
\caption{Quantization noise spectral slope versus bit depth for $\alpha = 2$ (20 trials, mean).}
\centering
\begin{tabular}{lll}
\toprule
Bits & Noise slope & Character \\
\midrule
4 & $-1.16$ & Strongly colored \\
5 & $-0.95$ & Colored \\
6 & $-0.32$ & Transitioning \\
7 & $-0.02$ & Approximately white \\
8 & $+0.00$ & White \\
\bottomrule
\end{tabular}
\label{tab:noise}
\end{table}

\begin{figure}[!t]
\centering
\begin{subfigure}[b]{0.48\columnwidth}
    \includegraphics[width=\textwidth]{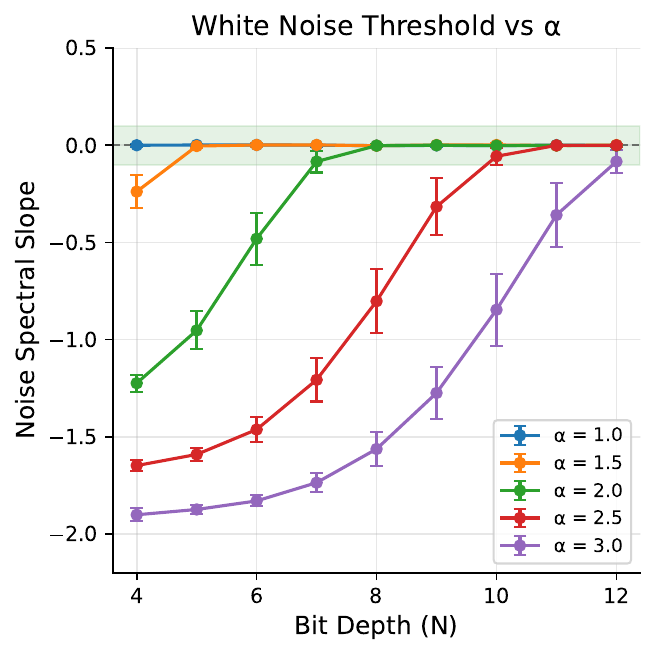}
    \caption{Threshold vs.\ $\alpha$}
\end{subfigure}
\hfill
\begin{subfigure}[b]{0.48\columnwidth}
    \includegraphics[width=\textwidth]{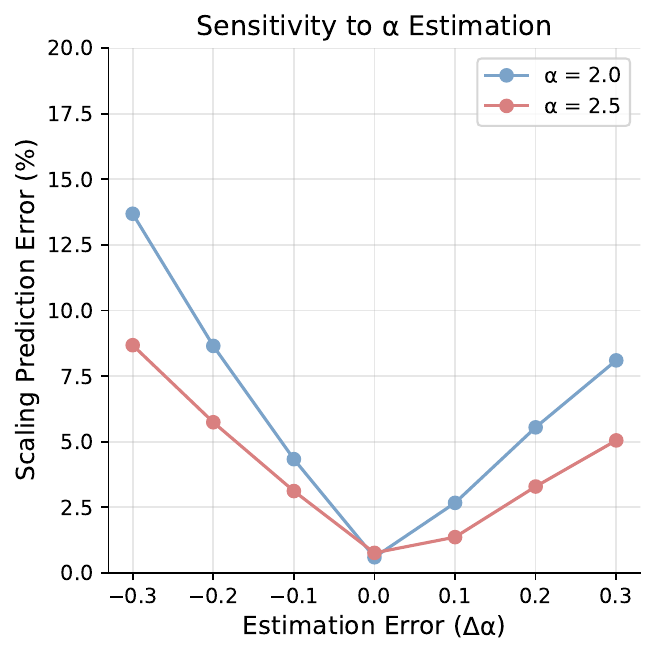}
    \caption{Sensitivity to $\alpha$}
\end{subfigure}
\caption{Robustness analysis. (a)~Quantization noise spectral slope versus bit depth for different $\alpha$; the green band marks $|$slope$|<0.1$ (approximately white). (b)~Scaling prediction error versus $\alpha$ estimation error.}
\label{fig:robustness}
\end{figure}

The transition point depends strongly on $\alpha$ (Fig.~\ref{fig:robustness}a). For $\alpha = 1$ (pink noise), the noise is already white at 4 bits, since the relatively uniform spectral energy ensures sufficient level crossings at all frequencies. For $\alpha = 2.5$, the threshold rises to 10 bits. For $\alpha = 3$, white noise behavior was not observed up to 12 bits. These results are consistent with the dithering literature \cite{roberts1962, schuchman1964, lipshitz1992, vanderkooy1987}, which developed techniques to whiten quantization noise precisely because it becomes colored at low bit depths.

\subsection{Scaling Law Validation}

We validated the scaling law on synthetic $1/f^\alpha$ signals for $\alpha \in \{1.5, 2.0, 2.5\}$ (20 trials each). The cutoff frequency at each bit depth was determined by finding where the signal PSD falls below the theoretical noise floor $\Delta^2 / (6 f_s)$. For $\alpha = 1.5$ we used $f_s = 200$~kHz and bit depths 5--6 (at higher bit depths the predicted $f_c$ exceeds the Nyquist frequency $f_s/2$). For $\alpha \in \{2.0, 2.5\}$ we used $f_s = 20$~kHz and bit depths 7--12.

The measured scaling ratios agree closely with predictions (Fig.~\ref{fig:scaling}): $2.55 \pm 0.05$ for $\alpha = 1.5$ (predicted 2.52, 1.3\% error), $1.99 \pm 0.03$ for $\alpha = 2.0$ (predicted 2.00, ${<}\,1\%$ error), and $1.74 \pm 0.03$ for $\alpha = 2.5$ (predicted 1.74, ${<}\,1\%$ error). When the noise floor is instead estimated empirically (by computing the PSD of the quantized signal and taking the median power in the upper quarter of the frequency range), the error increases to approximately 14\%, as spectral leakage from the dominant low-frequency components biases the estimated floor upward.

\subsection{Robustness}

\paragraph{Sensitivity to $\alpha$} In practice, $\alpha$ must be estimated from data. We evaluated sensitivity by computing the predicted scaling ratio at perturbed values of $\alpha$ and comparing against the true measured ratio (Fig.~\ref{fig:robustness}b). An estimation error of $\pm 0.1$ produces less than 5\% prediction error; $\pm 0.3$ yields up to ${\sim}\,14\%$. The sensitivity is slightly asymmetric, with underestimation of $\alpha$ producing larger errors.

\paragraph{Spectral peaks} Real $1/f^\alpha$ signals often contain narrowband peaks superimposed on the power-law background; electrophysiology data, for example, exhibits a prominent alpha peak near 10~Hz \cite{he2010, doyle2004}. We tested robustness by adding Gaussian spectral peaks of varying amplitude and center frequency to synthetic $1/f^2$ signals. Low-frequency peaks (10~Hz), even at $50\times$ the local $1/f^2$ level, produced less than 2\% error. A high-frequency peak at 100~Hz, near the cutoff region, increased the error to approximately 6\%.

\subsection{Application to Electrophysiology Data}

EEG data typically exhibits spectral slopes $\alpha \approx 1.5$--$1.6$ \cite{he2010, freeman2006}. The scaling law predicts that for $\alpha = 1.5$, each bit multiplies $f_c$ by $2.5\times$. At typical EEG sampling rates (160--256~Hz), the predicted $f_c$ exceeds the Nyquist frequency $f_s/2$ for $N \geq 5$ bits, leaving essentially no range at which the law is both valid ($N \geq N_{\min} = 5$) and testable below the Nyquist frequency.

\begin{figure}[H]
\centering
\begin{subfigure}[b]{0.48\columnwidth}
    \includegraphics[width=\textwidth]{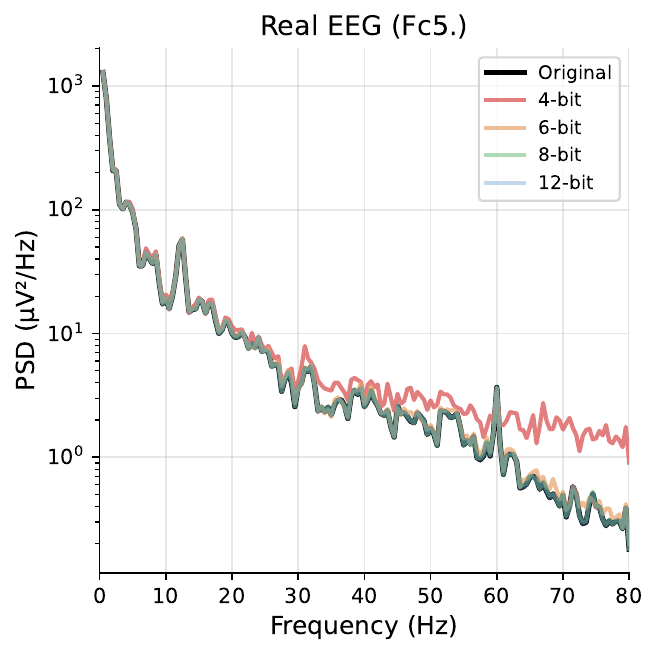}
    \caption{PSD at different bit depths}
\end{subfigure}
\hfill
\begin{subfigure}[b]{0.48\columnwidth}
    \includegraphics[width=\textwidth]{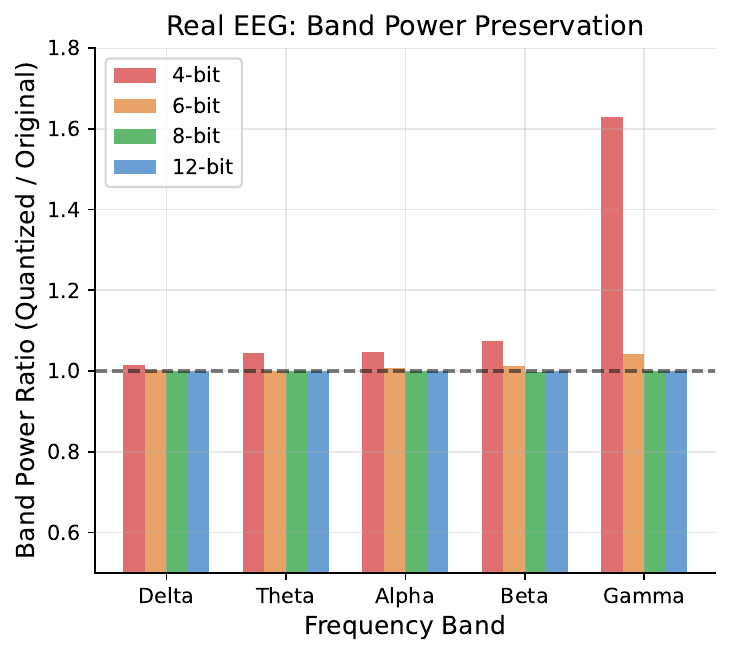}
    \caption{Band power preservation}
\end{subfigure}
\caption{EEG from PhysioNet \cite{physionet}. (a)~PSD showing the flat noise floor at 4 bits. (b)~Ratio of quantized to original band power. At 4 bits the Gamma band is distorted; at 6 bits and above all bands are preserved.}
\label{fig:eeg}
\end{figure}

To illustrate the practical effect, we applied the analysis to real electrophysiology data from the PhysioNet Motor Movement/Imagery dataset \cite{physionet} (160~Hz, $\alpha \approx 1.56$). Fig.~\ref{fig:eeg} shows that at 4 bits the Gamma band ($>$30~Hz) is severely distorted, while 6 bits suffice to preserve all standard bands. At 8 bits and above, $f_c$ exceeds the Nyquist frequency entirely. This provides concrete guidance for low-power wearable systems \cite{chi2010}, where each bit of ADC resolution directly impacts power consumption: reducing from 8 to 6 bits sacrifices no clinically relevant bandwidth at standard sampling rates.

\section{Discussion}

\subsection{The $\alpha$-Dependent Threshold}

The dependence of $N_{\min}$ on $\alpha$ has a physical interpretation. Signals with steep spectra concentrate their energy at low frequencies, so high-frequency components span only a few quantization levels even at moderate bit depths. This produces correlated quantization error that inherits the signal's spectral trend. For $\alpha \geq 3$, the white noise approximation may not hold at any practical bit depth, limiting the applicability of the scaling law to $\alpha \lesssim 2.5$ without dithering \cite{vanderkooy1987, lipshitz1992}.

\subsection{Relation to Prior Work}

The scaling law connects Bennett's SQNR formula \cite{bennett1948}, which establishes that each bit adds 6.02~dB of SNR, with the noise corner frequency concept where a colored process meets a white floor. While both components are classical, their combination into a bandwidth scaling law for power-law spectra is not present in the existing literature. The ADC and quantization theory literature \cite{bennett1948, widrow2008, gray1990} has focused on white or sinusoidal inputs, while the communities that routinely work with $1/f^\alpha$ signals (neuroscience, geophysics, audio) typically use high-resolution ADCs where quantization is not the limiting factor.

The result can be viewed as dual to the oversampling theorem \cite{candy1992}, which states that $4\times$ oversampling yields one additional bit of resolution for white noise. Our law instead characterizes how resolution translates to bandwidth for colored signals. The noise shaping literature \cite{lipshitz1992, candy1992} treats the inverse problem of deliberately redistributing quantization noise; our analysis quantifies the passive case. The observation that quantization noise becomes colored at low bit depths dates to the earliest PCM research \cite{clavier1947, widrow1956, oliver1948}; our contribution is the explicit bandwidth scaling law and the characterization of its $\alpha$-dependent validity threshold.

\subsection{Limitations}

The absolute cutoff (Eq.~\ref{eq:cutoff}) requires $S_0$, which is ill-defined for true $1/f$ processes whose PSD diverges as $f \to 0$ \cite{mandelbrot1968}: any finite-duration estimate of $S_0$ depends on the observation interval and does not converge to a fixed value, rendering the absolute cutoff impractical without a calibrated spectral reference. The scaling law (Eq.~\ref{eq:scaling}) cancels $S_0$ entirely, avoiding this problem. The law requires $N \geq N_{\min}(\alpha)$, restricting applicability for steep spectra. For shallow slopes ($\alpha < 2$), $f_c$ grows rapidly with $N$ (as $2^{2N/\alpha}$) and often exceeds the Nyquist frequency; for $\alpha = 1$ the cutoff surpasses the Nyquist frequency even at $N_{\min}$, precluding empirical validation without impractical sampling rates. The derivation assumes a pure power-law spectrum; while robust to low-frequency peaks ($<$2\% error), peaks near the cutoff degrade accuracy to ${\sim}\,6\%$. Extension to non-stationary signals or spectra with multiple regimes remains open.

\section{Conclusion}

We derived and validated a scaling law for the quantization-limited bandwidth of $1/f^\alpha$ signals: each additional bit of amplitude resolution extends the effective cutoff frequency by a factor of $2^{2/\alpha}$. The law holds when quantization noise is approximately white, requiring a minimum bit depth $N_{\min}$ that increases with $\alpha$. Validation on synthetic signals yields prediction errors below 3\%, and the result is robust to spectral peaks typical of real signals. Given a signal's spectral slope $\alpha$ and bit depth $N$, the law predicts both the frequency above which quantization noise dominates and the bandwidth gained or lost by changing resolution. For $\alpha = 2$ (Brownian-type spectra), each bit doubles $f_c$, so moving from 8-bit to 12-bit extends effective bandwidth by $16\times$; for $\alpha = 1.5$ (electrophysiology), each bit yields a $2.5\times$ increase. Conversely, the law identifies the minimum bit depth needed to preserve a frequency band of interest, informing ADC selection for any application involving power-law spectra.

\bibliographystyle{IEEEtran}
\bibliography{references}

@article{bennett1948,
  title={Spectra of quantized signals},
  author={Bennett, William Ralph},
  journal={The Bell System Technical Journal},
  volume={27},
  number={3},
  pages={446--472},
  year={1948},
  publisher={Nokia Bell Labs}
}

@article{gray1990,
  title={Quantization noise spectra},
  author={Gray, Robert M},
  journal={IEEE Transactions on Information Theory},
  volume={36},
  number={6},
  pages={1220--1244},
  year={1990},
  publisher={IEEE}
}

@book{widrow2008,
  title={Quantization noise: roundoff error in digital computation, signal processing, control, and communications},
  author={Widrow, Bernard and Kollar, Istvan},
  year={2008},
  publisher={Cambridge University Press}
}

@article{clavier1947,
  title={Distortion in a pulse count modulation system},
  author={Clavier, AG and Panter, PF and Grieg, DD},
  journal={Transactions of the American Institute of Electrical Engineers},
  volume={66},
  number={1},
  pages={989--1005},
  year={1947},
  publisher={IEEE}
}

@article{lipshitz1992,
  title={Quantization and dither: A theoretical survey},
  author={Lipshitz, Stanley P and Wannamaker, Robert A and Vanderkooy, John},
  journal={Journal of the Audio Engineering Society},
  volume={40},
  number={5},
  pages={355--375},
  year={1992}
}

@article{roberts1962,
  title={Picture coding using pseudo-random noise},
  author={Roberts, Lawrence G},
  journal={IRE Transactions on Information Theory},
  volume={8},
  number={2},
  pages={145--154},
  year={1962},
  publisher={IEEE}
}

@misc{physionet,
  title={{PhysioNet}: The {EEG} Motor Movement/Imagery Dataset},
  author={Goldberger, Ary L and Amaral, Luis AN and Glass, Leon and Hausdorff, Jeffrey M and Ivanov, Plamen Ch and Mark, Roger G and Mietus, Joseph E and Moody, George B and Peng, Chung-Kang and Stanley, H Eugene},
  year={2000},
  howpublished={\url{https://physionet.org/content/eegmmidb/1.0.0/}},
  note={Accessed: 2026}
}

@article{widrow1956,
  title={A study of rough amplitude quantization by means of {N}yquist sampling theory},
  author={Widrow, Bernard},
  journal={IRE Transactions on Circuit Theory},
  volume={3},
  number={4},
  pages={266--276},
  year={1956},
  publisher={IEEE}
}

@article{schuchman1964,
  title={Dither signals and their effect on quantization noise},
  author={Schuchman, Leonard},
  journal={IEEE Transactions on Communication Technology},
  volume={12},
  number={4},
  pages={162--165},
  year={1964},
  publisher={IEEE}
}

@article{jimenez2011,
  title={On the validity of the white noise hypothesis for quantization error analysis},
  author={Jimenez, Juan C and Shmaliy, Yuriy S and Ibarra-Manzano, Oscar},
  journal={SIAM Journal on Applied Mathematics},
  volume={71},
  number={6},
  pages={2165--2183},
  year={2011},
  publisher={SIAM}
}

@article{oliver1948,
  title={The philosophy of {PCM}},
  author={Oliver, Bernard M and Pierce, John R and Shannon, Claude E},
  journal={Proceedings of the IRE},
  volume={36},
  number={11},
  pages={1324--1331},
  year={1948},
  publisher={IEEE}
}

@article{sripad1977,
  title={A necessary and sufficient condition for quantization errors to be uniform and white},
  author={Sripad, Anand B and Snyder, Donald L},
  journal={IEEE Transactions on Acoustics, Speech, and Signal Processing},
  volume={25},
  number={5},
  pages={442--448},
  year={1977},
  publisher={IEEE}
}

@article{vanderkooy1987,
  title={Resolution below the least significant bit in digital systems with dither},
  author={Vanderkooy, John and Lipshitz, Stanley P},
  journal={Journal of the Audio Engineering Society},
  volume={35},
  number={3},
  pages={106--113},
  year={1987}
}

@book{candy1992,
  title={Oversampling delta-sigma data converters: theory, design, and simulation},
  author={Candy, James C and Temes, Gabor C},
  year={1992},
  publisher={IEEE Press}
}

@article{he2010,
  title={The temporal structures and functional significance of scale-free brain activity},
  author={He, Biyu J},
  journal={Neuron},
  volume={66},
  number={3},
  pages={353--369},
  year={2010},
  publisher={Elsevier}
}

@article{freeman2006,
  title={Origin, structure, and role of background {EEG} activity. Part 1. Analytic amplitude},
  author={Freeman, Walter J and Holmes, Mark D and Burke, Brian C and Vanhatalo, Sampsa},
  journal={Clinical Neurophysiology},
  volume={114},
  number={12},
  pages={2089--2107},
  year={2003},
  publisher={Elsevier}
}

@article{wenz1962,
  title={Acoustic ambient noise in the ocean: Spectra and sources},
  author={Wenz, Gordon M},
  journal={The Journal of the Acoustical Society of America},
  volume={34},
  number={12},
  pages={1936--1956},
  year={1962},
  publisher={Acoustical Society of America}
}

@book{aki1980,
  title={Quantitative Seismology: Theory and Methods},
  author={Aki, Keiiti and Richards, Paul G},
  year={1980},
  publisher={W.H. Freeman}
}

@article{mandelbrot1968,
  title={Fractional {B}rownian motions, fractional noises and applications},
  author={Mandelbrot, Benoit B and Van Ness, John W},
  journal={SIAM Review},
  volume={10},
  number={4},
  pages={422--437},
  year={1968},
  publisher={SIAM}
}

@article{chi2010,
  title={Dry-contact and noncontact biopotential electrodes: Methodological review},
  author={Chi, Yu Mike and Jung, Tzyy-Ping and Cauwenberghs, Gert},
  journal={IEEE Reviews in Biomedical Engineering},
  volume={3},
  pages={106--119},
  year={2010},
  publisher={IEEE}
}

@article{doyle2004,
  title={Quantification of the brain's {EEG} signal},
  author={Doyle, Orla M and Temko, Andriy and Marnane, William},
  journal={Digital Signal Processing},
  volume={14},
  number={6},
  pages={546--567},
  year={2004},
  publisher={Elsevier}
}

\end{document}